\documentstyle[12pt]{article}
\begin{document}
\newcommand{\siml}{\stackrel{<}{\sim}}
\newcommand{\simg}{\stackrel{>}{\sim}}
%
Running title: {\it Metal-Insulator Transition 
in Doubly Degenerate Hubbard's model}
\baselineskip=2\baselineskip
\begin{center}
{\large\bf The Metal-Insulator Transition }  \\
\end{center}
\begin{center}
{\large\bf  in the Doubly Degenerate Hubbard Model }  \\
\end{center}

\begin{center}
Hideo Hasegawa$^\dagger$  \\
{\it Department of Physics, Tokyo Gakugei University  \\
Koganei, Tokyo 184, Japan}
\end{center}
\begin{center}
{\rm (Received April 28, 1997)}
\end{center}
\thispagestyle{myheadings}
%
\begin{center} 
{\bf Abstract}   \par
\end{center} 

   A systematic study has been made on the metal-insulator (MI) 
transition of the doubly degenerate Hubbard model (DHM)
in the paramagnetic ground state, by using the slave-boson 
mean-field theory which is equivalent to the Gutzwiller 
approximation (GA).  For the case of infinite electron-electron 
interactions, we obtain the analytic solution, which becomes exact 
in the limit of infinite spatial dimension. On the contrary, 
the finite-interaction case is investigated by numerical 
methods using the simple-cubic model with the 
nearest-neighbor hopping.  The mass-enhancement factor, 
$Z$, is shown to increase divergently as one approaches 
the integer fillings ($N = 1, 2, 3$), at which the MI 
transition takes place, $N$ being the total number of electrons.
The calculated $N$ dependence of $Z$  is compared with
the observed specific-heat coefficient, 
$\gamma$, of ${\rm Sr}_{1-x}{\rm La}_{x}{\rm TiO}_3$ which is 
reported to significantly increase as $x$ approaches unity.

\vspace{0.5cm}
\vspace{0.5cm}

\noindent

\noindent
$\dagger$ e-mail address:  hasegawa@u-gakugei.ac.jp
\newpage
\noindent
{\large\bf $\S$1. Introduction}

  A study on the metal-insulator (MI) transition  has been
one of the most fascinating subjects in solid state physics.$^{1)}$
In recent years much attention has been paid to 
transition-metal compounds of
${\rm A}_{1-x}{\rm B}_{x}{\rm MO}_3$ where
A = Sr, Ca, B = La, Y, and M = Ti, V, Cr.$^{2,3)}$
These systems exhibit various phases such as 
a paramagnetic metal (PM),
paramagnetic insulator (PI) and
antiferromagnetic insulator (AFI) when the temperature ($T$),
pressure ($P$), magnetic field ($H$) 
or  chemical substitution ($x$) is changed.
Theoretically the MI transition has been so far investigated 
by using the single-band Hubbard model (SHM),$^{4-6)}$
with which observed $T-P$ and $T-x$ phase diagrams
have been {\it qualitatively} understood.$^{7-12)}$ 
We should, however, remind the fact that constituent transition 
metals of these systems inevitably have the orbital degeneracy.
Although the important role played by the orbital degeneracy 
has been recognized,                     
the additional degree of the orbital degeneracy has prevented us from
a systematic, theoretical analysis using the 
degenerate-band Hubbard model.

In the last few years,  several theoretical studies 
have been made of the MI transition in
the Hubbard model with orbital degeneracy,
by employing various methods such as
the Gutzwiller approximation (GA),$^{13-16)}$
Monte-Carlo method$^{17-19)}$ and  slave-boson theory.$^{20-22)}$
By using the GA, Lu$^{13)}$ obtained the critical interaction for
the MI transition, $U_c$, given by $U_c \sim (D + 1)$
at the half filling, 
while Gunnarsson {\it et al.}$^{17)}$
showed $U_c \sim \surd D$, using a diffusion Monte Carlo
method, $D$ denoting the orbital degeneracy. 
B\"{u}nemann and Weber$^{14)}$ discussed the 
first-order MI transition in the half-filled doubly degenerate
Hubbard model (DHM) when the exchange interaction is included
within the GA.
The MI transition of the infinite-dimensional DHM 
in the paramagnetic state is investigated by
Kotliar and Kajueter,$^{18)}$ and by Rozenberg$^{19)}$
using  the dynamical mean-field theory.

In a previous paper (referred to as I),$^{20)}$
the present author proposed the slave-boson functional-integral
theory for the degenerate-band Hubbard model, by adopting the
method originally employed for the Anderson lattice model.$^{23)}$ 
It is a generalization of the slave-boson method of the SHM$^{24)}$
to the degenerate-band model, and its  mean-field approximation is 
equivalent to the GA,$^{13-15,25)}$
which becomes exact for the Gutzwiller wavefunctions
in the limit of the infinite dimension.$^{16,26)}$
An alternative slave-boson method was independently
proposed by Fresard and Kotliar.$^{21)}$ 
In our second paper (referred to as II)$^{22)}$
the antiferromagnetic state of the half-filled DHM
was studied by using our slave-boson  theory.$^{20)}$

Although these studies$^{13-22)}$ are useful in understanding
the existing experimental data, it is more desirable to 
make a systematic study of the MI transition of the degenerate-band 
Hubbard model. We have made such a study by using the slave-boson
mean-field theory developed in I, which is
the purpose of the present paper.
In the next $\S$2,
we  briefly discuss the mean-field approximation to
our slave-boson theory$^{20)}$ applied to the DHM.
One of the advantages of our slave-boson mean-field theory over
sophisticated methods such as the dynamical 
mean-field method,$^{18,19)}$
is that we can obtain
analytic solutions in some limiting cases.
Indeed, we present, in $\S$3, the analytic solution
when the electron-electron interactions are infinite.
For finite interactions, we perform, in $\S$4, numerical
calculations using the simple-cubic model.
The final section ($\S$5) is devoted to conclusion and
discussion on the experimental data of 
${\rm Sr}_{1-x}{\rm La}_{x}{\rm TiO}_3$.$^{3)}$
\vspace{1.0cm}
\noindent
{\large\bf $\S$2. Formulation}

    
   We adopt the DHM whose Hamiltonian is given by
\begin{equation}
H = \sum_{\sigma} \sum_{i j} \sum_{m m'} t^{m m'}_{ij} 
c^\dagger_{im \sigma} c_{jm' \sigma}
+ \frac{1}{2} \sum_i \sum_{(m,\sigma) \neq (m',\sigma')}
U_{mm'}^{\sigma\sigma'}
c^\dagger_{im \sigma} c_{im \sigma}
c^\dagger_{im' \sigma'} c_{im' \sigma'},
\end{equation}
\noindent
where $c_{im\sigma}$ is an annihilation operator of an electron 
with an orbital index {\it m} 
and spin $\sigma (= \uparrow, \downarrow)$ on the lattice site $i$.
The electron hopping is assumed to be allowed only between the
same sub-band: $t_{ij}^{mm'} = t_{ij} \delta_{mm'}$ for
a simplicity.
The on-site interaction, $U_{mm'}^{\sigma \sigma'}$, is given by
\begin{eqnarray}
U_{mm'}^{\sigma\sigma'} &=& U_0 = U   \;\;\;\;\;\;\;\;\;\;
\mbox{for $m = m', \sigma \neq \sigma'$},  \\
&=& U_1 = U - 2J  \;\; 
\mbox{for $m \neq m', \sigma \neq \sigma'$},  \\
&=& U_2 = U - 3 J  \;\; 
\mbox{for $m \neq m', \sigma = \sigma'$}, 
\end{eqnarray}
where $U$ and $J$ are Coulomb and exchange interactions, respectively.

For the paramagnetic state of the DHM, we introduce the boson 
operators of $e_i$, $p_i$, $d_i$, $t_i$ and $f_i$  which denote
the empty, singly-, doubly-, triply- and fully-occupied states, 
respectively, at a given $i$ site.$^{20,22)}$
As for the doubly occupied states, we take into account 
three kinds of configurations;
$d_{i0}$ for a pair of electrons on the same 
orbital with opposite spin, 
$d_{i1}$ on the different  orbital with opposite spin, and
$d_{i2}$ on the different  orbital with same spin. 
These boson operators are under constraints of their completeness
and the correspondence with fermion operators.$^{20,22)}$

   When we apply our slave-boson mean-field theory developed in I
to the DHM under consideration, we get the ground-state energy
given by
\begin{eqnarray}
E = E_1 + E_2,
\end{eqnarray}
where
\begin{eqnarray}
E_1 & = & 2 \:q\: \varepsilon_0,    \nonumber  \\
E_2 & = & 2\:(U_0 d_0 + U_1 d_1 + U_2 d_2)
+ 2 \:(U_0 + U_1 + U_2) (2 t + f).  
\end{eqnarray}
Here the occupancies of $d_0$, $d_1$, $d_2$, $t$  and $f$ 
are the expectation values of respective boson operators 
and they are determined by the following self-consistent 
equations:$^{20,22)}$
\begin{equation}
U_0 + \varepsilon_0 \; (\partial q/\partial d_{0})  = 0,
\end{equation}
\begin{equation}
U_1 + \varepsilon_0 \;  (\partial q/\partial d_{1})  = 0,
\end{equation}
\begin{equation}
U_2 + \varepsilon_0\;  (\partial q/\partial d_{2})  = 0,
\end{equation}
\begin{equation}
2(U_0 + U_1 + U_2) +  \varepsilon_0\; (\partial q/\partial t)  = 0,
\end{equation}
\begin{equation}
(U_0 + U_1 + U_2) + \varepsilon_0\;  (\partial q/\partial f)  = 0.
\end{equation}
where  
\begin{equation}
q =  \frac{16  \left[ ( \surd \overline{d_{0}}
+  \surd \overline{d_{1}} + \surd \overline{d_{2}} )
(\surd \overline{p} + \surd \overline{t} ) + \surd \overline{e \, p} 
+ \surd \overline{t \, f}) \right]^2 }
{ N (4 - N)},
\end{equation}
\begin{equation}
\varepsilon_0 =  \int^{\mu} {\rm d}\,\varepsilon \; 
\sum_{\sigma} \; \varepsilon \; \rho_0(\varepsilon)
\end{equation}
\begin{equation}
N = 2 \;\int^{\mu} {\rm d}\,\varepsilon \; 
\sum_{\sigma} \; \rho_0(\varepsilon)
\end{equation}
with
\begin{equation}
e = 1 - N + 2 \; (d_{0} + d_{1} + d_{2}) + 8 t + 3 f,
\end{equation}
\begin{equation}
p = N/4  - (d_{0} + d_{1} + d_{2}) - 3 t -  f.
\end{equation}
In eqs. (7)-(16), $\mu$ is the Fermi level,
$\rho_0(\varepsilon)$ denotes the non-interacting
density of states per sub-band per spin, and $N$ is
the {\it total} number of electrons which is 1 and 2
for the quarter- and half-filled bands, respectively.
The band-narrowing factor, $q$, is expressed in terms of
various occupancies as given by eq. (12). 
%
The self-consistent equations given by eqs. (7)-(16) are equivalent
to those obtained by employing  the GA to
Gutzwiller's wavefunctions.$^{13-15)}$

\vspace{1.0cm}
\noindent
{\large\bf $\S$3. Analytic Solutions for Infinite Interactions}

When the interactions are infinite,
we can obtain analytic solutions
for the self-consistent equations given by eqs. (7)-(16).
Solutions for the less-than-half filling ($N \leq 2$)
are classified into the three cases 
depending on the value of $j$ defined by
\begin{equation}
j = \lim_{U \rightarrow \infty} \; (J/U).
\end{equation}
Solutions for the more-than-half filling ($N > 2$)  can be 
obtained from the result for $N \leq 2$ 
with the electron-hole symmetry.

\vspace{0.5cm}
\noindent
{\it 3.1 Case of $j = 0$}

When $j = 0$, three, double occupancies become equivalent:
$d_0 = d_1 = d_2 = d$.
Taking the $U \rightarrow \infty$ limit 
of eq. (7), (8) or (9) with
\begin{equation}
t = f = 0,
\end{equation}
we get
\begin{eqnarray}
d_0 = d_1 = d_2 &=& 0  
\;\;\;\;\;\;\;\;\;\;\;
\;\;\;\;\;\;\;\;\;\;\;
\;\;\;\;\;\;\;\;\;  \mbox{(for $ N \leq 1$)},  \\ 
&=& (1/6)\; ( N - 1 ) \;\;\;\;\;\;\;\;\;\; \mbox{(for  $1 < N < 2$)}.
\end{eqnarray}
From eqs. (18)-(20), $e$ and $p$ in eqs. (15) and (16) are
given by
\begin{eqnarray}
e &=& 1 - N  \;\;\;\;\;\;\;\;  \mbox{(for $ N \leq 1$)},  \\ 
&=& 0 \;\;\;\;\;\;\;\;\;\;\;\;  \mbox{(for  $1 < N \leq 2$)}.
\end{eqnarray}
\begin{eqnarray}
p &=& N/4  
\;\;\;\;\;\;\;\;\;\;\;\;\;\;\;  
\;\;\;\;\;\;\;\;\;  
\mbox{(for $ N \leq 1$)},  \\ 
&=& (1/4) \; (2 - N) \;\;\;\;\;\;  \mbox{(for  $1 < N \leq 2$)}.
\end{eqnarray}
Substituting eqs. (18)-(24) to eq. (12), 
we obtain the band-narrowing factor:
\begin{eqnarray}
q &=& 4\;(1 - N)/(4 - N)  
\;\;\;\;\;\;\;\;\;\;\;
\;\;\;\;\;\;\;\;\;\;\;
\mbox{(for $ N \leq 1$)},  \\ 
&=& 6 \; (N - 1)(2 - N)/N (4 - N)
\;\;\;\;\;\;\;  \mbox{(for  $1 < N \leq 2$)}.
\end{eqnarray}
Near $N = 1$ and 2, $q$ is given by
\begin{eqnarray}
q &=& (4/3) \; \delta \;\;\;\;\;\;\;\;  \mbox{(for $ N \siml 1$)},  \\ 
&=& 2 \; \delta \;\;\;\;\;\;\;\;\;\;\;\;\;  \mbox{(for $ N \simg 1$)},  \\ 
&=& (3/2) \; \delta \;\;\;\;\;\;\;\;\;  \mbox{(for  $ N \siml 2$)},
\end{eqnarray}
where the electron or hole doping concentration, $\delta$,
is given by $\delta = \; \mid 1 - N \mid$ 
in eqs. (27) and (28)
and $\delta = \; \mid 2 - N \mid$ in eq. (29).

\vspace{0.5cm}
\noindent
{\it 3.2 Case of $j > 0$ }

Similarly we get 
\begin{equation}
d_0 = d_1 = t = f = 0,
\end{equation}
\begin{eqnarray}
d_2 &=& 0  
\;\;\;\;\;\;\;\;\;\;\;\;\;\;  
\;\;\;\;\;\;\;\;\;\;\;\;\;\;  
\mbox{(for $ N \leq 1$)},  \\ 
&=& (1/2) \; (N - 1) \;\;\;\;\;\;\;  \mbox{(for  $1 < N \leq 2$)}.
\end{eqnarray}
The $N$-dependence of $e$ and $p$ is given by eqs. (21)-(24).
The band-narrowing factor becomes
\begin{eqnarray}
q &=& 4\;(1 - N)/(4 - N)  
\;\;\;\;\;\;\;\;\;\;\;\;\;
\;\;\;\;\;\;\;\;\;\;\;\;\;
\mbox{(for $ N \leq 1$)},  \\ 
&=& 2 \; (N - 1)(2 - N)/N (4 - N)
\;\;\;\;\;\;\;\;  \mbox{(for  $1 < N \leq 2$)}.
\end{eqnarray}
It is easy to see that $q$ close to $N = 1$ and 2 is given by
\begin{eqnarray}
q &=& (4/3) \; \delta \;\;\;\;\;\;\;\;  \mbox{(for $ N \siml 1$)},  \\ 
&=& (2/3) \; \delta \;\;\;\;\;\;\;\;  \mbox{(for $ N \simg 1$)},  \\ 
&=& (1/2)\;  \delta \;\;\;\;\;\;\;\;\;\;\;\;  \mbox{(for  $ N \siml 2$)},
\end{eqnarray}
where $\delta = \; \mid 1 - N \mid$ in eqs. (35) and (36)
and $\delta = \; \mid 2 - N \mid$ in eq. (37).

\vspace{0.5cm}
\noindent
{\it 3.3 Case of $j < 0$ }

We get$^{27)}$ 
\begin{equation}
d_1 = d_2 = t = f = 0,
\end{equation}
\begin{eqnarray}
d_0 &=& 0  
\;\;\;\;\;\;\;\;\;\;\;\;\;\;  
\;\;\;\;\;\;\;\;\;\;\;\;\;  
\mbox{(for $ N \leq 1$)},  \\ 
&=& (1/2) \;  (N  - 1) \;\;\;\;\;\;\;  \mbox{(for  $1 < N \leq 2$)}.
\end{eqnarray}
The $N$-dependences of $e$, $p$ and $q$ are same as
in the $j > 0$ case.

It is worth to remark that in the SHM, the solution for
$U \rightarrow \infty$ with $N \leq 1$ is given by
\begin{equation}
e = 1 - N,
\end{equation}
\begin{equation}
p = N/2,
\end{equation}
with $d = 0$, leading to
\begin{eqnarray}
q &=& 2\;(1 - N)/(2 - N)  \\ 
&\sim& 2 \;\delta
\;\;\;\;\;\;\;\;\;\;\;  
\;\;\;\;\;\;\;\;\;\;\;  
\;\;\;\;\;\;\;\;\;\;\;  
\mbox{(for  $N \siml 1$)},
\end{eqnarray}
where $\delta = \mid 1 - N \mid$.
We should note that these solutions of the GA
for infinite interactions
discussed above become exact in the limit of the infinite 
spatial dimension.$^{16,26)}$ 
Some figures depicting the $N$ dependence of 
the occupancies and the band-narrowing 
factor for the DHM will be shown shortly 
(dashed curves in Figs. 3 - 5).

\vspace{1.0cm}
\noindent
{\large\bf $\S$4. Numerical Calculations for Finite Interactions}

When the interactions are finite, we cannot obtain the
analytic solution for the self-consistent equations 
given by eqs. (7)-(16), which have to be solved
by numerical methods for given parameters of $N$, $U$ and $J$
and non-interacting density of states, $\rho_0(\varepsilon)$. 
We adopt the simple-cubic model band,
expressed  by an approximate, analytic form:$^{28)}$ 
\begin{eqnarray}
\rho_0(\varepsilon) &=& A \left[ 9 - \omega^2 \right]^{1/2}
- C \left[ 1 - \omega^2 \right]^{1/2}  
\;\;\:\:\:\:\: 
\mbox{ for $\mid \omega \mid \le 1$}, \nonumber \\
&=& A \left[ 9 - \omega^2 \right]^{1/2}
- B \left[ 1 - ( \mid \omega \mid - 2)^2 \right]^{1/2},  
\;\; \mbox{ for $1 <  \mid \omega \mid \le 3$},  \nonumber \\
&=& 0  \;\;\;\;\;\;\;\;\;\;\;\;\;\;\;\;\;
\mbox{for $\mid \omega \mid > 3$},
\end{eqnarray}
with $\omega = \varepsilon/2t$, $A/2t$ = 0.101081, 
$B/2t= 0.128067$ and $C/2t = 0.02$, $t$ being the nearest-neighbor
hopping.
The energy and the interactions are hereafter measured in units
of a half of the total band width ($W/2 = 6 t = 1$).

\vspace{0.5cm}
\noindent
{\it 4.1 Results as a function of $U$}

Firstly we discuss the $U$ dependence of  the band-narrowing
factor, $q$, at the integer fillings of $N = 1$ and 2.$^{29)}$
Figure 1(a)  shows the $U$ dependence of $q$ 
for various values of $J/U$ of the quarter-filled case 
$(N = 1)$.
When the $U$ is increased, the $q$ factor decreases and
vanishes above $U_c$ which stands for the critical
interaction for the MI transition.
We get $q \sim (U_c - U)$ at $U \leq U_c$, which is the same
as in the SHM.$^{30)}$
In all the  cases shown in Fig. 1(a), the MI transition is of the
second order.
The calculated $U_C$ is plotted by the solid curve
in Fig. 2 as a function of $J/U$,
which shows an increase in $U_c$  as increasing $J/U$.
On the contrary, the behavior of the MI transition in the half-filled 
band ($N = 2$) is quite different from that in the $N = 1$ case, 
as is shown in Fig. 1(b).
The MI transition becomes the {\it first-order} one 
for finite $J/U$,$^{14,20)}$
and  $U_c$ {\it decreases} as increasing the 
absolute value of $J/U$.  Then the $U_c$ curve has
a cusp at $J/U = 0$, as shown by the dashed curve in Fig. 2.

\vspace{0.5cm}
\noindent
{\it 4.2 Results as a function of $N$}

Figures 3(a)  shows the $N$ dependence of
the occupancies for various $U$ values with $J/U = 0$.
At $N = 0$ only the empty occupancy $e$ is finite ($e = 1$).
As increasing $N$ from zero, $e$ monotonously decreases while  $p$ has a
peak at $N = 1$, where the double occupancies,
$d_0 = d_1 = d_2 = d$, begin 
to increase. Since $t$ and $f$ are one or two-order magnitude
smaller than the other occupancies, they are not shown 
in the Figure.
Dashed curves express analytic solutions of  
$e$, $p$ and $d$ for $U = \infty $ 
given by eqs. (18)-(24). 
We note that as the $U$ value is increased, 
the $N$ dependence of the
occupancies approaches that for $U = \infty$ 
and that the occupancies for $U = 4$ are almost the same as those for 
$U = \infty$.

Similarly, the $N$ dependence of the occupancies for 
various $U$ values with $J/U = 0.1$ is shown in Fig. 3(b).
General behavior  for $J/U = 0.1$ is very similar to that 
for $J/U = 0$, except that only $d_2$ has an appreciable 
magnitude among  the three, double occupancies
because $U_0 > U_1 > U_2$ for $J > 0$.

Figures 4(a) and 4(b) show  $q$
as a function of $N$ for various
$U$ with $J/U =0$ and $J/U = 0.1$, respectively.
The band-narrowing factor is unity at $N = 0$, and
it has dips or zeros at $N = 1$ and 2.
Our result of the $N$ dependence of $q$ for $J/U = 0$ is 
similar to those reported in refs. 19 and 21.
The  mass-enhancement factor, $Z$, 
which is given as the inverse of $q$, 
for $J/U =0$ and $J/U = 0.1$ is shown in Figs. 5(a) and 5(b),
respectively, as a function of $N$.
For intermediate interactions ($U \sim W$),  
$Z$ has  peaks at the integer
band fillings of $N = 1$ and 2, where $Z$ shows the 
divergence for the strong interaction ($U \gg W$).
As increasing the $U$ value, $Z$ increases as expected.
Dashed curves in Figs. 5(a) and 5(b) express the results for 
$U = \infty$ given by eqs. (25)-(26) and (33)-(34).
We note that   $Z$ at $N \leq 1$  is
not symmetric with that at $N > 1$ 
and it is not the same as that
at $N \leq 2$, in particular for $J/U = 0.1$.
This is more clearly seen in the case of $U = \infty $
as shown by dashed curves in Figs. 5(a) and 5(b) and 
as given by eqs. (27)-(29) and (35)-(37),
which show the $\delta^{-1}$ divergence  as approaching 
the integer fillings.

\vspace{1.0cm}
\noindent
{\large\bf $\S$5. Conclusion and Discussion}

By using the slave-boson mean-field theory$^{20)}$ 
which is equivalent to the GA,$^{13-16,25)}$
we have systematically studied the DHM, which shows 
the MI transition to take place at the integer fillings 
of $N = 1, 2$ and 3 in the paramagnetic state.
When the magnetically ordered state is taken into account, 
this MI  transition is modified, particularly
in the half-filling case, where the antiferromagnetic 
state can be realized for a relatively small interaction.
Indeed, for the simple-cubic model with the nearest-neighbor hopping,
the antiferromagnetic insulator (AFI)
is stabilized for an infinitesimally small interaction,
and then the MI transition does not take
place except at the vanishing interaction point
in the DHM,$^{22)}$ just as in the SHM.$^{24,31)}$ 
The schematic, phase diagram of the MI transition
of the DHM is shown in Fig. 6.
When approaching the integer filling from below or above
with the  moderate interaction ($U \sim W$),
the mass-enhancement factor, $Z$, is considerably 
increased at the PM-PI transition. On the contrary,
when the PM-AFI (PM-FI) transition occurs with the 
strong interaction and/or the favorable band filling
(as the half filling), 
the mass-enhancement is not realized in the AFI (FI)
because of the presence of
the energy gap in the density of states,$^{32)}$
although the enhancement of $Z$ 
is expected as approaching this transition from the metallic side.
Our calculations
applying the slave-boson mean-field theory to the DHM,$^{20,22)}$
have well accounted for the essential features 
of the MI transition of a system with orbital degeneracy:

    It has been reported that the linear coefficient of the
low-temperature specific heat, $\gamma$, and the susceptibility
of ${\rm Sr}_{1-x}{\rm La}_{x}{\rm TiO}_3$ significantly increase 
as $x$ approaches unity.$^{3)}$
Filled circles in Figs. 5(a) and 5(b) express the observed 
mass-enhancement ratio of $\gamma/\gamma_0 \; (= Z)$. 
We note that the calculated $N$ dependence of $Z$ 
well reproduces the observed $x$ dependence of 
$\gamma/\gamma_0$.
It was previously tried to explain the experimental data 
with the use of the SHM,$^{12)}$ 
in which the mass-enhancement 
factor, $Z$, of its paramagnetic state 
divergently increases  near the half filling 
(eqs. (43) and (44)).
We should, however, remind the fact that 
this  enhancement does not occur if the antiferromagnetic
state is included, as was discussed above (Fig. 6). 
For a better understanding of the MI transition in 
${\rm Sr}_{1-x}{\rm La}_{x}{\rm TiO}_3$,
we have to adopt the triply degenerate Hubbard model 
in a strict sense, because  $d$ electrons in 
this system belong to  $t_{2g}$ bands.$^{3)}$ 
In order to discuss the high-energy phenomena
of transition-metal compounds$^{2,3)}$
such as photoemission and the optical conductivity,
it is necessary to take into account 
"incoherent-quasiparticle" states
arising from boson fluctuations around 
the saddle-point solution:$^{33)}$
our slave-boson mean-field theory
includes only "coherent-quasiparticle" states, 
which are relevant to the low-energy phenomena like 
the mass enhancement.
Going beyond the mean-field approximation is left as one of
our future subjects in the slave-boson functional integral 
theory for the degenerate-band Hubbard model.


\noindent

\vspace{1.0cm}
\noindent{\bf Acknowledgment}  \par
\vspace{0.2cm}

This work is partly supported by
a Grant-in-Aid for Scientific Research from the Japanese 
Ministry of Education, Science and Culture.

\newpage
\newpage

\newpage
\noindent
{\large\bf  Figure Captions}   \par
\vspace{1.0cm}
\noindent
{\bf Fig. 1} 
The $U$ dependence of the band-narrowing factor, $q$,
for various $J/U$ with (a) $N = 1$ and (b) $N = 2$. 
\vspace{0.5cm}

\noindent
{\bf Fig. 2} 
The critical interaction for the MI transition,
$U_c$, in the paramagnetic state  
as a function of $J/U$ for $N = 1$ and 2,
the solid (dashed) curve denoting the second (first)
order transition.
\vspace{0.5cm}

\noindent
{\bf Fig. 3} 
The $N$ dependence of the occupancies
for $U = 2$ (thin solid curve), 4 (bold solid curve) 
and $\infty$ (dashed
curve) with (a) $J/U = 0$ and (b) $J/U = 0.1$.
\vspace{0.5cm}

\noindent
{\bf Fig. 4} 
The $N$ dependence of  the
band narrowing factor, $q$, for various $U$
with (a) $J/U = 0$ and (b) $J/U = 0.1$,
the dashed curve denoting the result for
$U = \infty$. 
\vspace{0.5cm}

\noindent
{\bf Fig. 5} 
The $N$ dependence of  the
mass-enhancement factor, $Z (=1/q)$, for various $U$
with (a) $J/U = 0$ and (b) $J/U = 0.1$,
the dashed curve denoting the result for
$U = \infty$. 
Circles express the experimental data of the
enhancement ratio of $\gamma/\gamma_0$ in
${\rm Sr}_{1-x}{\rm La}_{x}{\rm TiO}_3$ [ref. 3],
$\gamma_0$ being taken to be the $\gamma$ value at 
$N = 0$.
\vspace{0.5cm}

\noindent
{\bf Fig. 6} 
The electron number vs. interaction phase diagram 
of DHM (schematic),
showing the paramagnetic metal (PM), paramagnetic insulator (PI),
antiferromagnetic insulator (AFI) and ferromagnetic
insulator (FI).   The relative stability between AFI and FI 
at the quarter filling depends on the electronic 
and lattice structures and
on the interaction  of a given system.


\begin{thebibliography}{99}
\bibitem{Saurenbach} Although the exchange interaction $J$ is conventionally
believed to be positive, we include the negative-$J$ case 
for a completeness of our discussion.
\bibitem{Saurenbach}Some calculated results of the $U$ dependence
of the occupancies in the paramagnetic DHM have been reported in ref. 20.
\end{thebibliography}
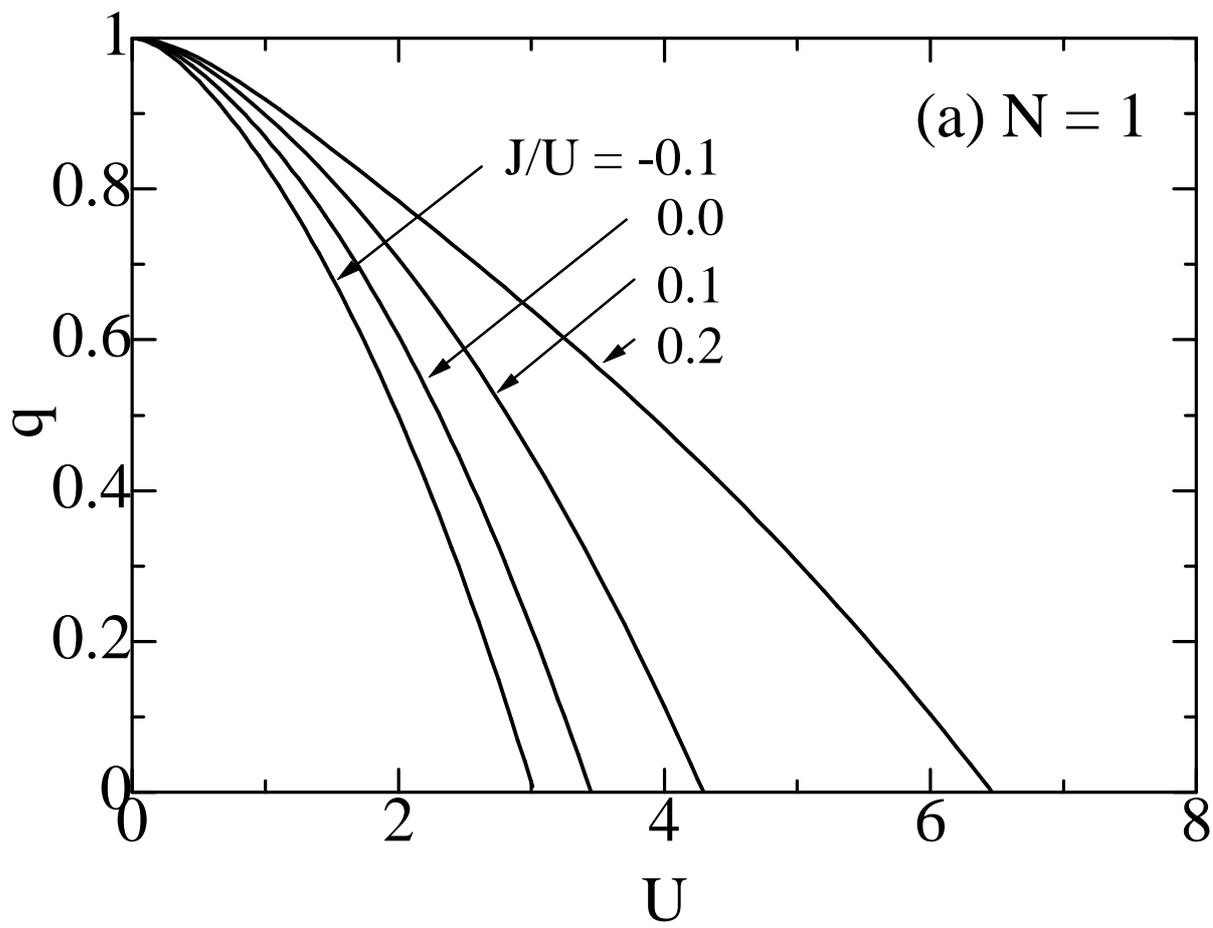
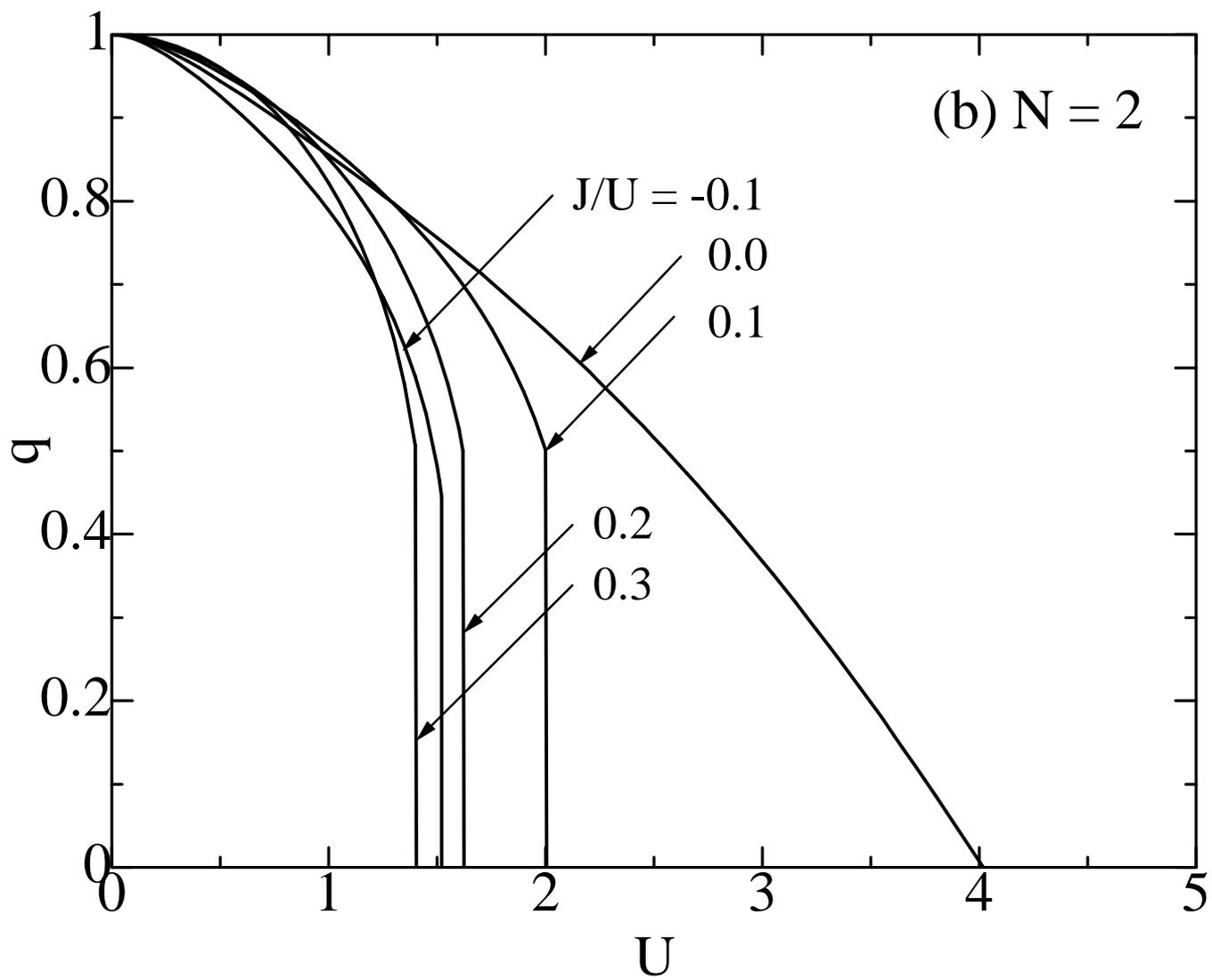
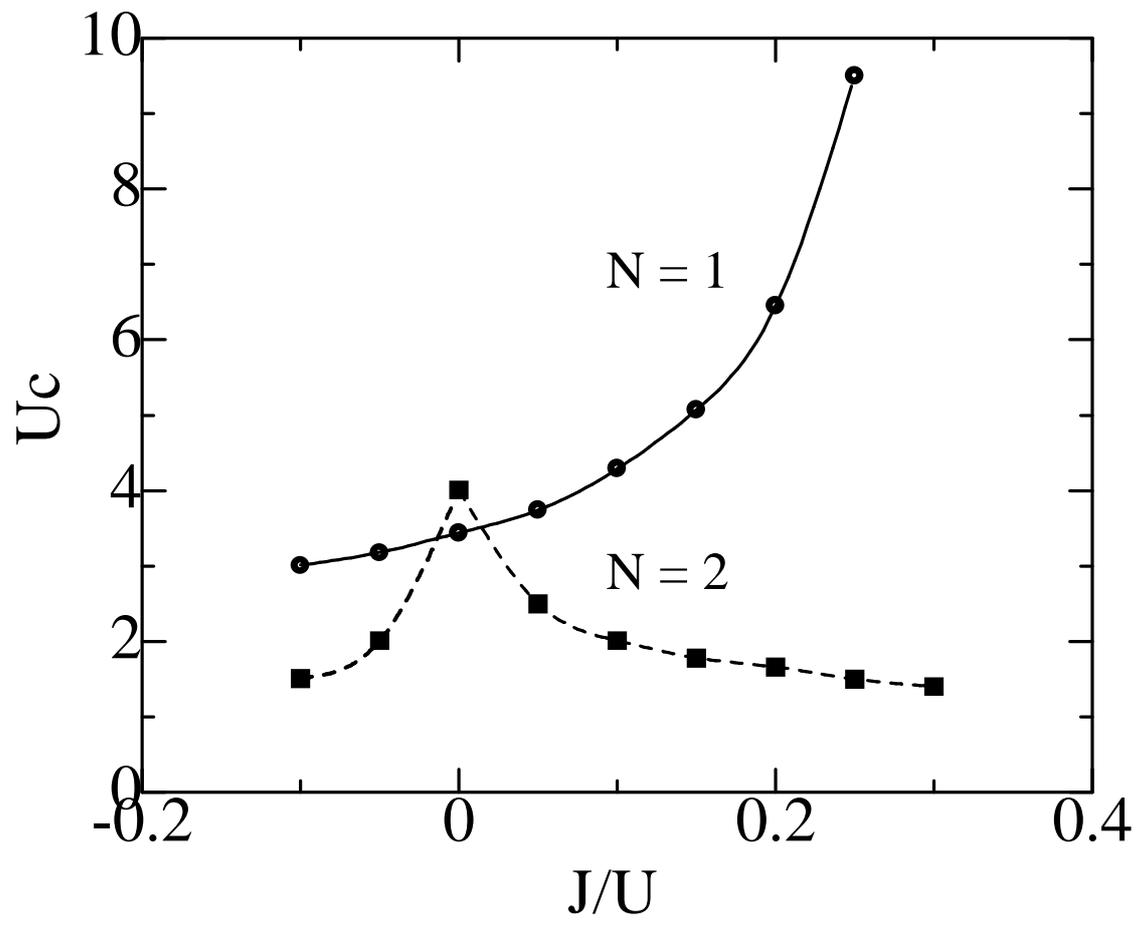
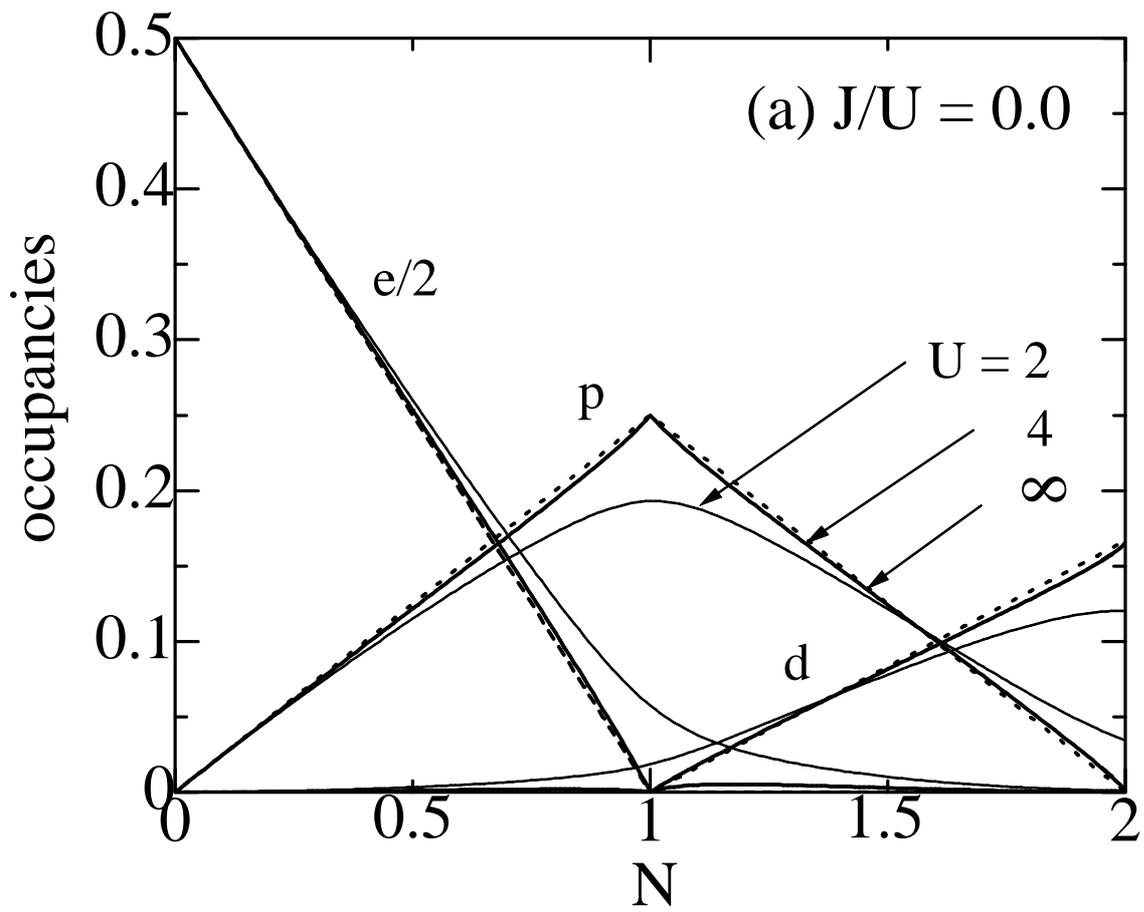
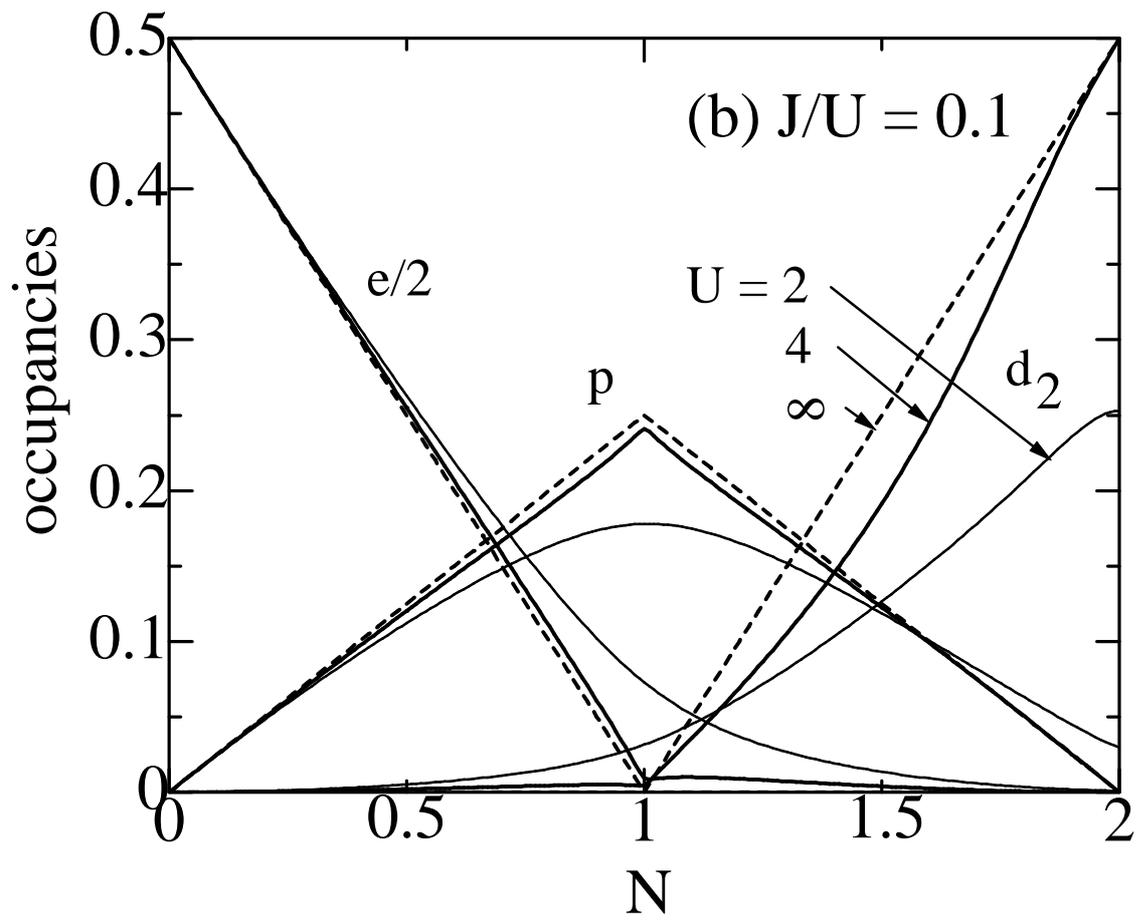
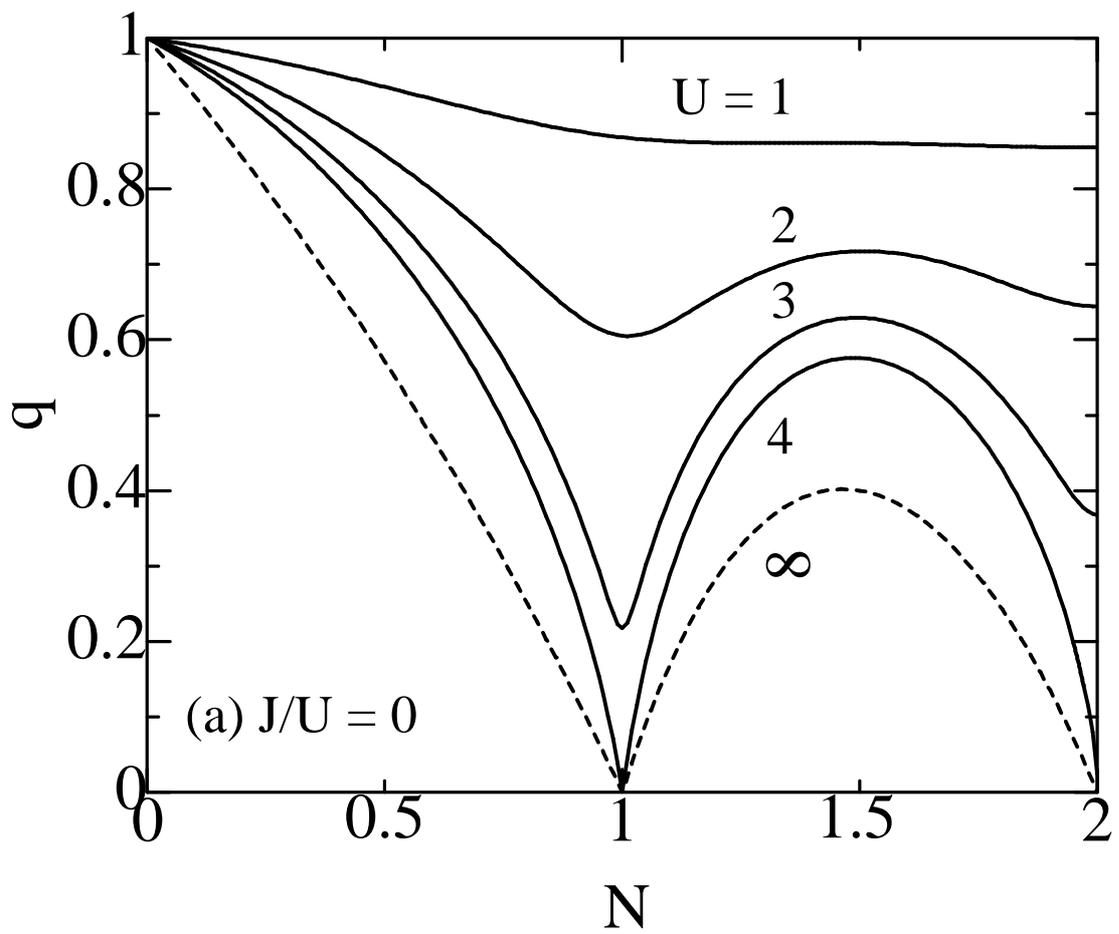
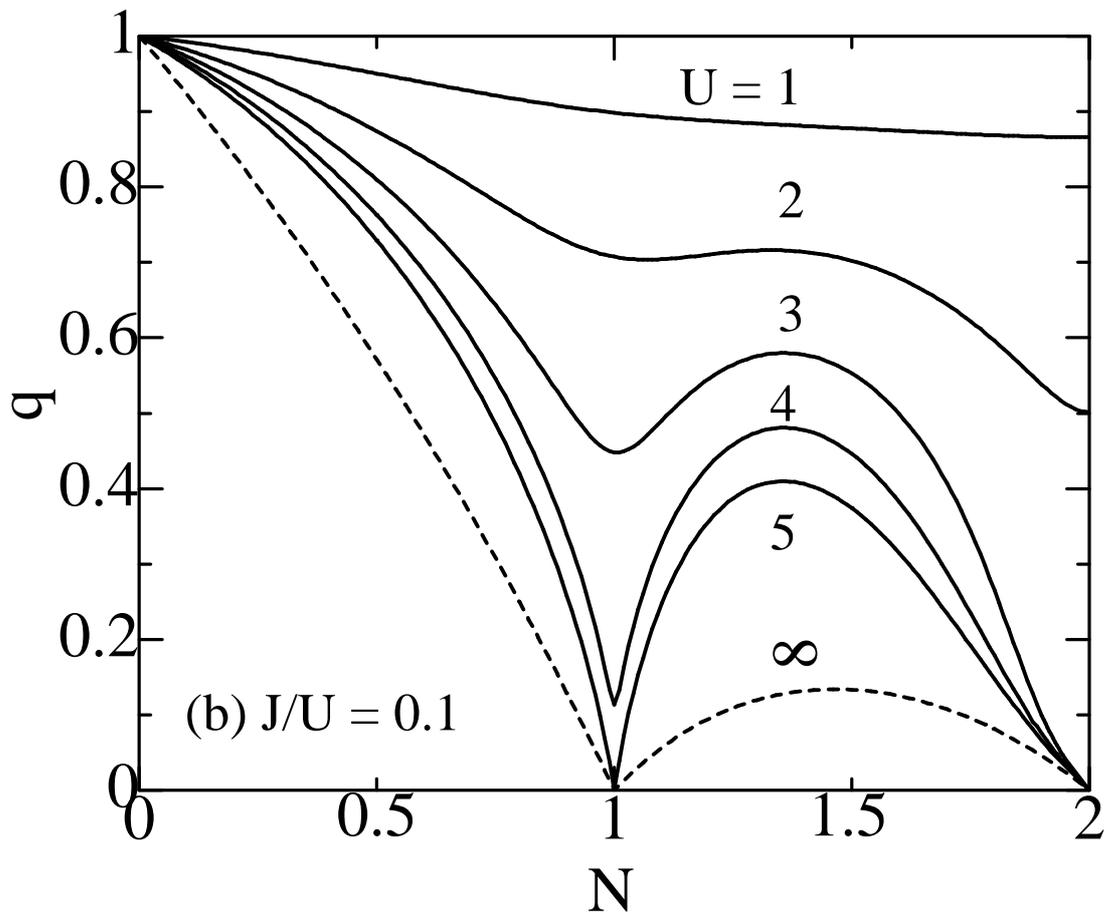
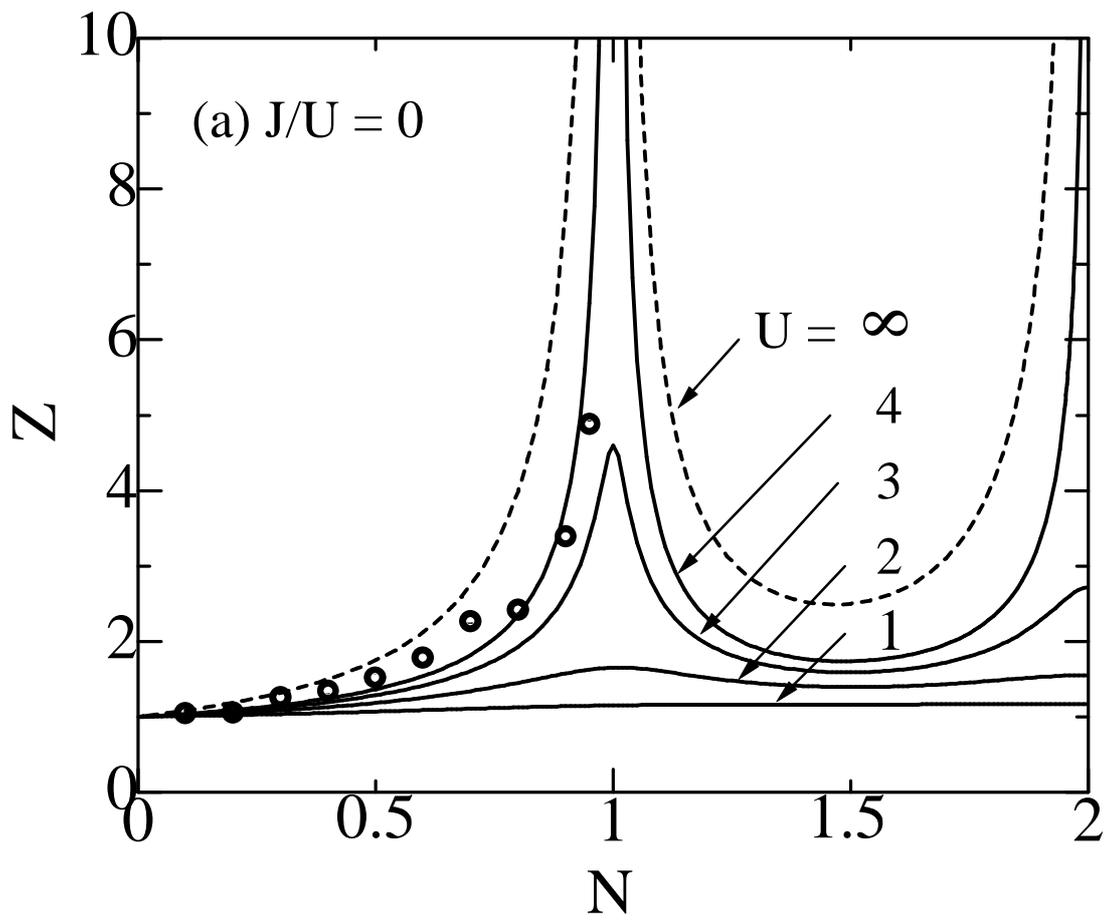
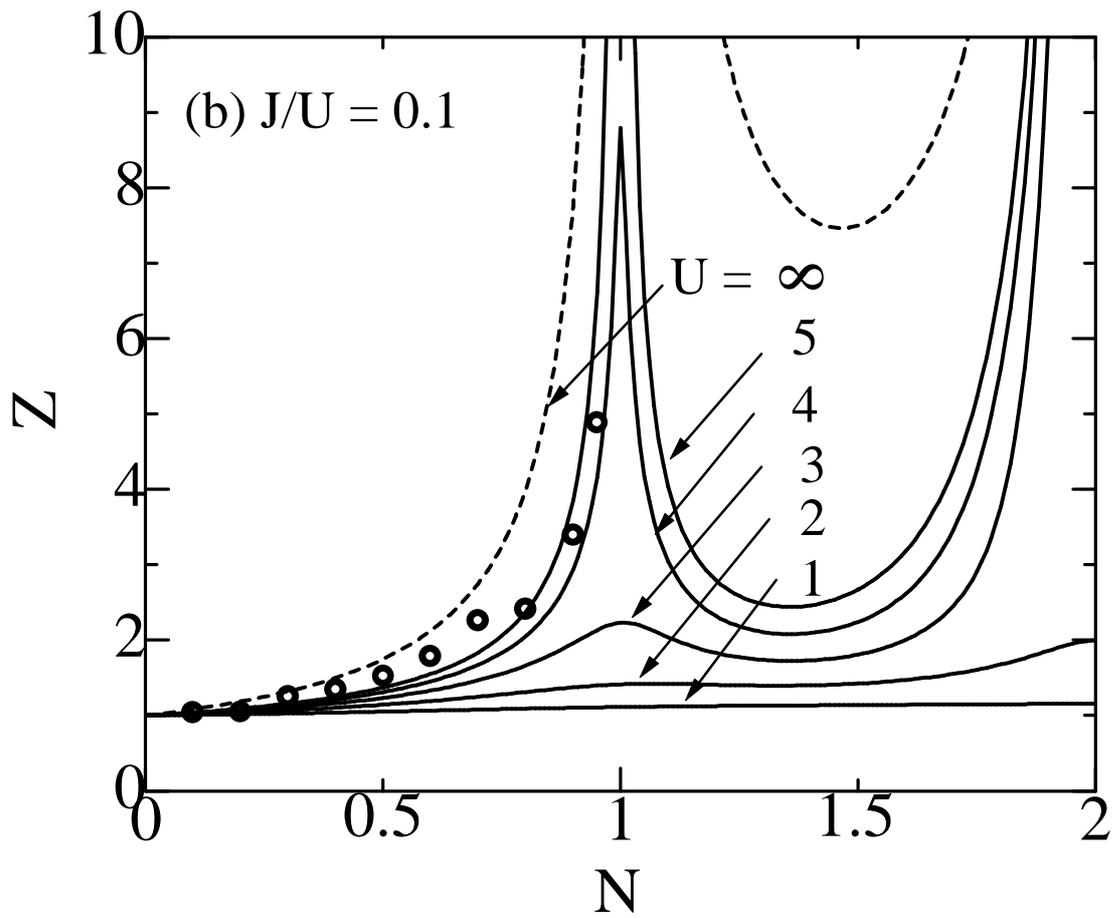
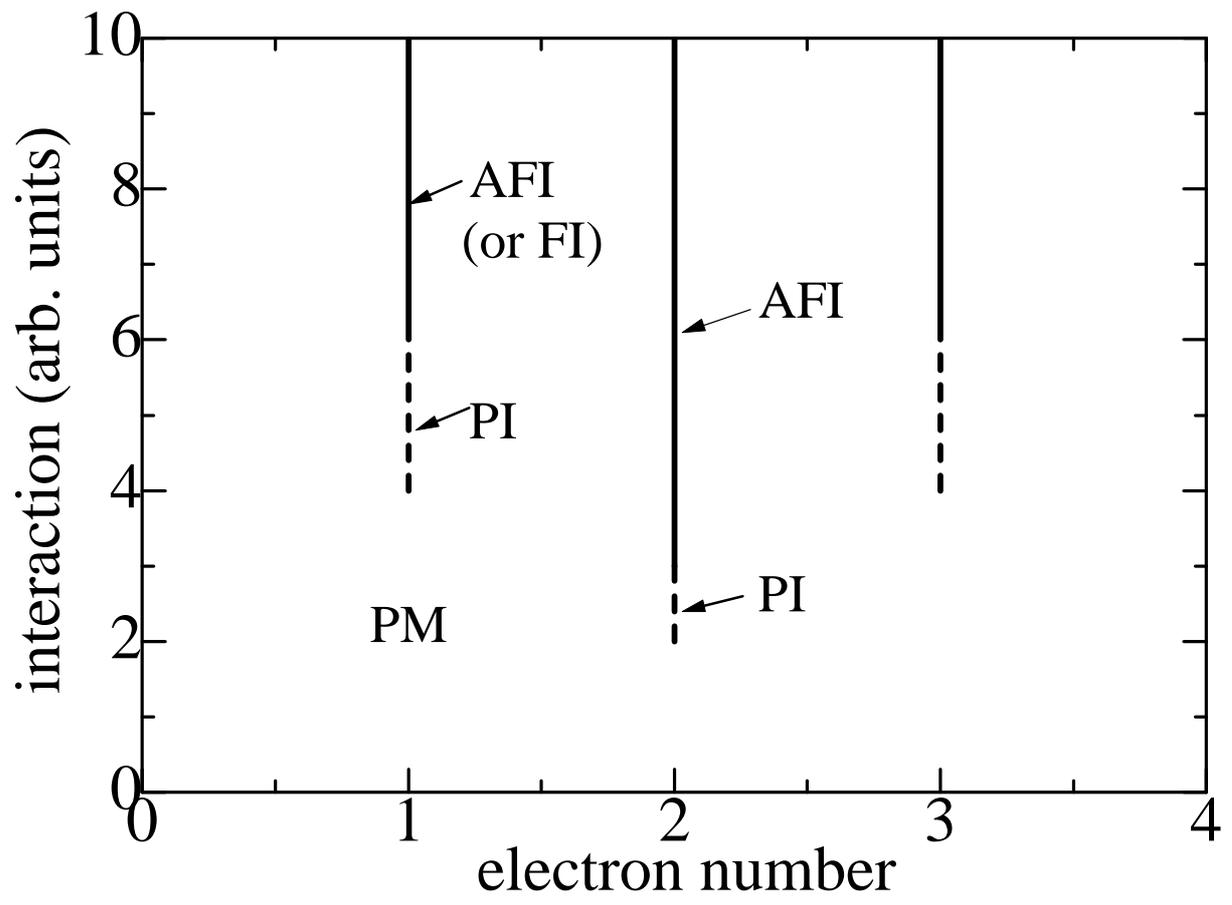
\end{document}